\documentclass[10pt,superscriptaddress,aps,pra,twocolumn,showpacs,floatfix]{revtex4-2}
\usepackage[utf8]{inputenc}
\usepackage{graphicx,amsmath,amsfonts,amssymb}
\usepackage{color}
\usepackage[colorlinks=true, allcolors={blue}]{hyperref}
\usepackage{graphicx}
\usepackage{epstopdf}
\usepackage{float}
\usepackage{placeins}
\usepackage{lipsum}
\usepackage{mathtools}
\usepackage{hyperref}
\usepackage{comment}

\usepackage{soul}

\usepackage{silence}
\newcommand{\orcid}[1]{\href{https://orcid.org/#1}{\includegraphics[width=7pt]{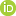}}}

\usepackage{xcolor}

\usepackage[normalem]{ulem}

\usepackage{eso-pic}

\newcommand{\EditorsSuggestionText}{\large\scshape Editors' Suggestion}

\AddToShipoutPictureBG*{%
  \ifnum\value{page}=1\relax
    \AtTextUpperLeft{%
      \hspace{0pt}
      \raisebox{1.2\baselineskip}[0pt][0pt]{\EditorsSuggestionText}
    }%
  \fi
}
\begin{document}

\preprint{APS/123-QED}

\title{Temperature and non-Markovian parameter estimation in quantum Brownian motion}

\author{João C. P. Porto~\orcid{0009-0006-6639-1413}}
\email{carlosciaufpi@gmail.com}
\affiliation{Departamento de F\'{i}sica, Universidade Federal do Piau\'{i}, Campus Ministro Petr\^{o}nio Portela, CEP 64049-550, Teresina, PI, Brazil}

\author{Carlos H. S. Vieira~\orcid{0000-0001-7809-6215}}
\email{vieira.carlos@ufabc.edu.br}
\affiliation{Centro de Ci\^{e}ncias Naturais e Humanas, Universidade Federal do ABC,
Avenida dos Estados 5001, 09210-580 Santo Andr\'e, S\~{a}o Paulo, Brazil.}

\author{Irismar G. da Paz~\orcid{0000-0002-9613-9642}}
\affiliation{Departamento de F\'{i}sica, Universidade Federal do Piau\'{i}, Campus Ministro Petr\^{o}nio Portela, CEP 64049-550, Teresina, PI, Brazil}

\author{Pedro R. Dieguez~\orcid{0000-0002-8286-2645}}
\affiliation{International Centre for Theory of Quantum Technologies, University of Gdańsk, Jana Bażyńskiego 1A, 80-309 Gdańsk, Poland}

\author{Lucas S. Marinho~\orcid{0000-0002-2923-587X}}
\email{lucas.marinho@ufpi.edu.br}
\affiliation{Departamento de F\'{i}sica, Universidade Federal do Piau\'{i}, Campus Ministro Petr\^{o}nio Portela, CEP 64049-550, Teresina, PI, Brazil}


\begin{abstract}
We investigate a quantum metrological protocol operating in a non-Markovian environment by employing the quantum Brownian motion (QBM) model, in which the system is linearly coupled to a reservoir of harmonic oscillators. Specifically, we use a position-momentum (PM) correlated Gaussian state as a probe to examine how memory effects influence the evolution of the system's covariance matrix in the weak coupling regime under both high- and low-temperature conditions. To confirm the presence of non-Markovian behavior, we apply two well-established non-Markovianity quantifiers. Furthermore, we estimate both the channel's sample temperature and its non-Markovianity witness parameter. Our results demonstrate that non-Markovianity and PM correlations can jointly be valuable resources to enhance metrological performance.
\end{abstract}

\maketitle


\section{Introduction}\label{sec:intro}

Brownian motion originally described the random movement of heavy particles suspended in a liquid or gas, driven by collisions with the constituents of the surrounding fluid \cite{Brown1928}. This phenomenon played a key role in supporting the atomist hypothesis of matter \cite{Einstein1905}. It became a fundamental model in physics, characterized by its non-deterministic nature and the presence of dissipative processes due to unavoidable interactions with the environment \cite{MazoBokk2008,lampo2019BrownianMotion,caldeira1983a,caldeira1983b,JPPaz1992PRD,Schlosshauer2,IlluminatiPRA2018}. Quantum Brownian Motion (QBM) translates these ideas into the quantum domain and has become essential in the study of open quantum systems \cite{Petruccione2002book,weiss2008book}. For example, it is the first realistic decoherence model used to study the dynamics of quantum Darwinism \cite{ZurekPRL2008,ManiscalcoSciRep2016}. Since the early days of quantum mechanics, significant research has been dedicated to understanding the quantum dynamics of a small system interacting with its surroundings, with QBM serving as a key framework in this area \cite{Petruccione2002book}, resulting in applications such as quantum communication~\cite{VasilePRA2011}, thermodynamics~\cite{WuPRE2023}, and metrology~\cite{Mirkhalaf_2024NJP}. 

Quantum metrology, in particular, focuses on identifying the most effective methods that employ quantum resources to perform high-resolution and highly sensitive measurements of physical parameters, surpassing the limits achievable in classical physics~\cite{LuPRA2010}, known as the shot-noise limit (SNL). Applications of quantum metrology are numerous and include: superresolution imaging~\cite{BraunPRA2023}; ultra-precise clocks~\cite{RosenbandSCIENCE2008}; navigation systems~\cite{GracePhysRevApplied2020}; magnetometry~\cite{Budker2007NATURE}; optical and gravitational-wave interferometry~\cite{DEMKOWICZDOBRZANSKI2015,TsePRL2019}; and thermometry~\cite{WengPRL2014,WuPRR2021,AnJunHongPRApplied2022,LandiPRA2024}. In the context of QBM, recent studies have investigated quantum thermometry with nonclassical probes~\cite{Mirkhalaf_2024NJP}. Also, an enhancement of low-temperature thermometry by strong coupling using a Brownian thermometer was proposed~\cite{LCorreaPRA2017}.

The Markovian approximation involves neglecting short-time system-reservoir correlations that arise due to the structure of the reservoir spectrum~\cite{VasilePRA2009}. In this context, the Born approximation is typically assumed, advocating for weak coupling between the system and environment~\cite{Petruccione2002book}. However, in scenarios with strong system-environment couplings, structured and finite reservoirs, low-temperatures, and considerable initial system-environment correlations, non-Markovian evolution must be taken into account~\cite{Liu2011NatPhys}. Nonetheless, even in the weak coupling limit~\cite{VasilePRA2009} and/or high-temperature regime~\cite{IlluminatiPRA2018}, there are instances of a structured environment where sufficiently long system-reservoir correlations necessitate a non-Markovian treatment. Recent advancements in experimental techniques and novel materials have enabled the exploration of regimes where non-Markovian effects are significant~\cite{ParisPRA2011,Liu2011NatPhys,ParisPRA2014,Groblacher2015Nature}. While Markovian processes erase information and gradually diminish the distinguishability of initial states~\cite{LuPRA2010}, non-Markovian dynamics preserve memory effects, unlocking new opportunities.

Experiments have shown that, contrary to common assumptions in high-temperature quantum Brownian motion, the spectral density can exhibit strongly non-Ohmic behavior, leading to pronounced non-Markovian dynamics~\cite{Groblacher2015Nature}. Additionally, the transition from Markovian to non-Markovian dynamics in open quantum systems was realized with optical systems~\cite{Liu2011NatPhys}, enabling applications such as continuous-variable quantum key distribution, where non-Markovian properties of lossy channels can be exploited to enhance security~\cite{VasilePRA2011}. Moreover, it was demonstrated that the resulting non-Markovian dynamics allow quantum correlated states to surpass metrological strategies that rely on uncorrelated states, even when employing otherwise identical resources~\cite{PlenioPRL2012}, which is in marked
contrast with general Markovian noise, where an arbitrarily small amount
of noise is enough to restore the scaling dictated by the standard quantum limit.

However, the application of quantum metrology to non-Markovian processes has received considerably less attention~\cite{YangPRL2021}. For instance, Gaussian quantum thermometry employing nonclassical probes within the QBM framework has been explored, yet these investigations have been restricted to the Markovian regime~\cite{Mirkhalaf_2024NJP}.
Nevertheless, non-Markovian dynamics have been shown to enhance thermometer performance~\cite{AllatiPRE224} and play a crucial role in achieving optimal precision in noisy quantum optical metrology~\cite{AnPRL2019}. Although a quantum Brownian probe in the non-Markovian regime can enhance thermometry in the low-temperature regime~\cite{LCorreaPRA2017}, the analysis was limited only to the steady-state regime where the non-Markovian features are less relevant. Building on these insights, we explore in this work the metrology of QBM in the non-Markovian regime within the transient temporal domain, where memory effects are more prominent. Our results show that non-Markovian effects can provide a metrological advantage over scenarios traditionally studied within the Born-Markov approximation. To investigate this, we adopt the well-established QBM model, known for its exact solvability~\cite{VasilePRA2009}. We analyze the dynamics of quantum metrology in noisy continuous variable (CV) quantum systems and investigate the effects of non-Markovian memory on the estimated quantities. Additionally, we explore the use of a parameterized Gaussian probe, which may or may not exhibit initial position-momentum (PM) correlations~\cite{OzielMPLA2019,LustosaPRA2020,MarinhoPRA2020,Marinho2024SciRep,Porto2024}, as a resource for metrology. By comparing correlated Gaussian states with standard ones, we find that PM correlations can enhance the non-Markovian effect and improve metrology. 

\begin{figure}[ht]
\centering
\includegraphics[scale = 0.59]{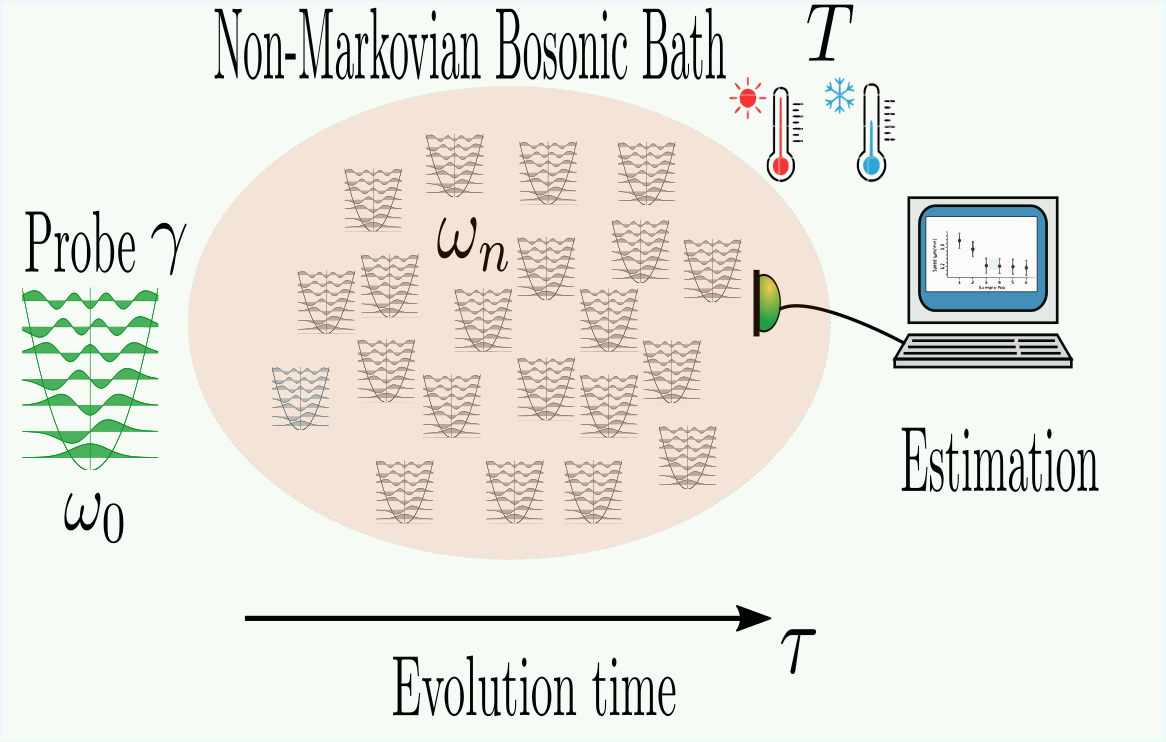}
\caption{Estimation protocol scheme: An initial probe state with frequency $\omega_0$ is prepared, exhibiting a non-zero position-momentum correlation $(\gamma \neq 0)$. The probe undergoes non-Markovian dynamics during evolution through a bosonic quantum Brownian motion channel at temperature $T$. In this stage, the environmental parameters of the system are encoded into the probe state via non-unitary evolution. Subsequently, measurements are made to estimate the unknown channel parameters. The non-Markovian regime is controlled by adjusting the ratio $x=\omega_c/\omega_0$
 , which compares the cut-off frequency $\omega_c$ of the bath to the frequency $\omega_0$ of the probe state. }
\label{fig1}
\end{figure}

This work is structured as follows. In Sec.~\ref{sec: tf}, we present the theoretical framework, including the non-Markovian QBM model (see Fig.~\ref{fig1}) and quantum Fisher information. We discuss our main results in Sec. \ref{sec:re}, showing how non-Markovian effects and position-momentum correlations can enhance QBM metrology.  Finally, in Sec.~\ref{sec:disc}, we draw our main conclusions and perspectives.

\section{THEORETICAL FRAMEWORK}~\label{sec: tf}

This section introduces the main framework for examining the metrology advantage via the evolution within a non-Markovian Gaussian channel.

\subsection{Gaussian states and QBM}

A state $\rho$ in continuous-variable systems comprising $N$ bosonic modes is represented by its corresponding characteristic function $\chi (\rho)[\boldsymbol{r}] = \text{Tr}[\rho D(\boldsymbol{r})]$, and via Fourier transform to a Wigner function $W(\boldsymbol{r})=\frac{1}{(2\pi)^{2N}}\int d^{2N}\boldsymbol{\widetilde{r}}\exp{[-i\boldsymbol{\widetilde{r}}^{\text{T}}\boldsymbol{\Omega} \boldsymbol{r} ]} \chi (\boldsymbol{\widetilde{r}})$, with $D(\boldsymbol{r})= \exp{[i\boldsymbol{r}^{\text{T}}\boldsymbol{\Omega} \boldsymbol{r} ]}$ the displacement operator, $\boldsymbol{r} = (\hat{q}_1,\hat{p}_1,\dots,\hat{q}_N,\hat{p}_N)^{\text{T}}$ the quadrature vector that satisfies the commutation relation $[\boldsymbol{r}_i,\boldsymbol{r}_j]=i\hbar \boldsymbol{\Omega}_{ij}$. Here, $\boldsymbol{ \Omega} = \bigoplus_{k=1}^{N} \boldsymbol{\omega}$ is the symplectic matrix, with $\boldsymbol{ \omega}=i\sigma_{y}$, where $\sigma_{y}$ the Pauli matrix~\cite{RevModPhys2012Lloyd}. Any quantum state $\rho$ is regarded as Gaussian if its characteristic function and Wigner representation are Gaussian
\begin{equation}
    \chi (\boldsymbol{r}) = \exp\left[-\frac{1}{2}\boldsymbol{r}^{\text{T}}(\boldsymbol{\Omega}\boldsymbol{\sigma}\boldsymbol{\Omega}^{\text{T}})\boldsymbol{r}-i(\boldsymbol{\Omega d}^{\text{T}})\boldsymbol{r}\right].
\end{equation}

\begin{equation}
  W (\boldsymbol{r}) =\frac{ \exp\left[-\frac{1}{2}(\boldsymbol{r}-\boldsymbol{d})^{\text{T}}\boldsymbol{\sigma}^{-1}(\boldsymbol{r}-\boldsymbol{d})\right]}{(2\pi)^N \sqrt{\text{det}\boldsymbol{\sigma}}}.
\end{equation}
Consequently, this state can be uniquely described by its first (the displacement vector $\boldsymbol{d}$) and its second (covariance matrix $\boldsymbol{\sigma}$) moments, respectively
\begin{equation}
\boldsymbol{d} = \langle \boldsymbol{r} \rangle,
\end{equation}
\begin{equation}
\boldsymbol{\sigma}_{ij} = \frac{1}{2} \langle \{ \Delta\hat{r}_i, \Delta\hat{r}_j  \}\rangle,
\end{equation}
where $ \Delta\hat{r}_i = \hat{r}_i -\langle \hat{r}_i\rangle$, and $\{\hat{A}, \hat{B}\}$ represent the anticommutator between the operators $\hat{A}$ and $\hat{B}$.

A Gaussian quantum channel is an operation that preserves the Gaussianity of a Gaussian input state \cite{PolzikBook2007}. In our analysis, we consider a dynamic 
one-mode quantum harmonic oscillator of frequency $\omega_0$ interacting with $N$ independent bosonic quantum oscillators of the environment, which correspond to the well-stabilized Quantum Brownian Motion (QBM)~(See Fig. \ref{fig1}). The exact master equation, derived without relying on the Born-Markov approximation or any other simplifying assumptions, is given by~\cite{JPPaz1992PRD,IlluminatiPRA2018}
\begin{gather}
 \frac{d \rho (t)}{dt}  = - \frac{i}{\hbar} [H_S,\rho (t)] - \Delta(t) [\hat{q}, [\hat{q},\rho (t)]] + \Pi(t) [\hat{q}, [\hat{p},\rho (t)]] \nonumber \\  - i\Gamma(t) [\hat{q}, [\hat{p},\rho (t)]], \label{eq:master_QBM}
\end{gather}
where $\Gamma(t)$ is the damping coefficient, $\Delta(t)$ and  $\Pi(t)$ are the normal and anomalous diffusion coefficients, respectively (see Appendix~\ref{Ap:QBM}). The dynamics of QBM can be described through a map defined by the following transformation in the displacement vector $\boldsymbol{d}$ and the covariance matrix $\boldsymbol{\sigma}$ \cite{VasilePRA2009,IlluminatiPRA2018,PolzikBook2007}
\begin{equation}
\boldsymbol{d}(t) = \mathcal{S}(t) \boldsymbol{d}(0),  
\end{equation}
\begin{equation}
\boldsymbol{\sigma} (t) = \mathcal{S}(t) \boldsymbol{\sigma}(0) \mathcal{S}^{\mathsf{T}}(t) +\mathcal{T}(t),  
\end{equation}
where $ \mathcal{S}(t)$ and $ \mathcal{T}(t)$ are $2N \times 2N$ matrices defined in Appendix~\ref{Ap:QBM}. Physically, $\mathcal{S}(t)$ describes ampliﬁcation, attenuation, and rotation in phase
space, whereas the $ \mathcal{T}(t)$ contribution is a noise term. For instance, in standard Brownian motion, temperature $T$ influences the dynamics solely through the noise matrix $\mathcal{T}(t)$, while the drift matrix $\mathcal{S}(t)$ remains independent of temperature \cite{Mirkhalaf_2024NJP}. The explicit form of these matrices depends on the system-environment coupling, which is characterized by the spectral density function $J(\omega)$, where $\omega$ represents the bath oscillator's frequency (see Appendix \ref{Ap:QBM}). The coupling between the system and the environment is modeled by the spectral density, which characterizes the properties of the bath. In order to obtain analytical expressions for the damping coefficient $\Gamma(t)$, the normal and anomalous diffusion coefficients, $\Delta(t)$ and $\Pi(t)$, in Eq. (\ref{eq:master_QBM}), it is common in the literature to assume that the system is linearly coupled (in frequency) to a continuum of harmonic oscillators \cite{Petruccione2002book,BreuerPRA2020}. Here, we consider an Ohmic-like class of spectral density distributions
\begin{equation}\label{eq:spectral_density}
    J_s(\omega) =  \Big ( \frac{\omega}{\omega_c} \Big )^s \;e^{  - \omega/\omega_c },
\end{equation}
with the parameter $s$ characterizing the system-environment properties. The case with $s=1$ is known as Ohmic, and for $s < 1$ ($s > 1$) we have a sub-Ohmic (super-Ohmic) case, respectively. The cutoff frequency of the environment, $\omega_c$, limits the contribution of environmental frequencies with infinite values. In this work, we will focus on the Ohmic case. 

Recently, correlations between two conjugated operators have been used to enhance Gaussian quantum metrology~\cite{Porto2024}. To highlight the advantages of non-Markovian dynamics in quantum metrology and the role of information backflow from the environment to the system, we also explore here a position-momentum (PM) correlated state~\cite{DODONOV1980PLA} as the initial state. This state can be characterized by zero first moments $\boldsymbol{d}=0$, and the covariance matrix 
\begin{equation}\label{eq:iniial_state}
    \boldsymbol{\sigma}(0) = \frac{1}{2}\begin{pmatrix} \;
1 & \gamma \\
\gamma & 1+\gamma^2 \;
\end{pmatrix},
\end{equation}
assuming a non-null initial position-momentum correlation ($ \boldsymbol{\sigma}_{12}=\boldsymbol{\sigma}_{21}=\gamma/2$). The real parameter $\gamma$ can take values in the interval $-\infty < \gamma < \infty$ and ensures the correlation for the initial state. The correlated Gaussian state was introduced in Ref.~\cite{DODONOV1980PLA}, and it is unitary equivalent to the squeezed states with complex
squeezing parameters~\cite{dodonov2002nonclassical}. By examining the Pearson correlation coefficient between the quadratures $\hat{q}$ and $\hat{p}$, given by $\mathcal{P}=\sigma_{12}/\sqrt{\sigma_{11}\sigma_{22}}\,(-1\leq \mathcal{P} \leq1)$, the parameter $\gamma$ is derived as $\gamma=\mathcal{P}/\sqrt{1-\mathcal{P}^{2}}\,(-\infty\leq \gamma\leq\infty)$. Therefore, $\gamma$ represents a parameter that encodes the initial correlations between $\hat{q}$ and $\hat{p}$ in the initial state. The specific case $\gamma=0$ corresponds to a simple, uncorrelated, coherent Gaussian state with uncertainties identical to those of the vacuum state obeying the SNL~\cite{Marinho2024SciRep}. Recently, a similar state was employed for thermometry, but they disregarded phase space rotations~\cite{LandiPRA2024}, which is equivalent to setting $\gamma=0$. It is worth noting that although the most general squeezed state may exhibit PM correlations, not every correlated state is squeezed. In this context, our focus is on exploring how these correlations lead to metrological gains rather than investigating the squeezing phenomenon itself, although there may be connections between these effects \cite{MarinhoPRA2020,Marinho2024SciRep} (see Appendix \ref{Ap:PM_metrology}).

\subsection{Non-Markovianity}

Here, we review a non-Markovianity criterion and its corresponding measure, which is based on the violation of the divisibility property of a dynamical map. This approach directly assesses the channel structure without relying on the evolution of specific quantum states~\cite{IlluminatiPRL2015,IlluminatiPRA2018}.  
A dynamical map $\mathcal{E}$ is a general quantum operation that transforms a quantum state, defined as $\mathcal{E}: \rho \rightarrow \mathcal{E}(\rho)$, while preserving the trace, i.e., $\text{Tr}[\mathcal{E}(\rho)] = 1$. This quantity is a more robust quantifier because state-based measures of non-Markovianity, which rely on the distinguishability of quantum states, may fail in certain cases. This limitation arises because such methods provide only a sufficient—but not necessary—condition for identifying non-Markovian behavior \cite{Rivas_2014}.  
Moreover, assessing the divisibility violation of the dynamical map does not require an extensive maximization over all quantum states to detect nonmonotonicity in state distances \cite{IlluminatiPRA2018,BreuerPRA2020}. In addition, in our results section, we will introduce a state-based quantifier for comparative analysis—specifically, the purity—and examine the conditions under which it accurately reflects non-Markovian behavior, as well as the limitations beyond which it ceases to be a reliable indicator.

We consider the following set of trace-preserving linear maps $\{ \mathcal{E} (t_2,t_0), t_2 \geq t_0  \}$, representing the system's evolution from time $t_0$ to $t_2$. The map is said to be
divisible, or Markovian, if, for every $t_2$ and $t_1$, it holds that $\mathcal{E} (t_2,t_0) = \mathcal{E} (t_2,t_1) \mathcal{E} (t_1,t_0), \;  t_2 \geq t_1 \geq t_0,$ where the map $\mathcal{E} (t_2,t_1)$  is completely positive (CP)~\cite{IlluminatiPRL2015}. An evolution is considered non-Markovian if it does not satisfy this divisibility property. As demonstrated in Ref.~\cite{IlluminatiPRA2018} for the case of the QBM channel, the system exhibits non-Markovian evolution whenever the non-Markovianity quantifier
\begin{equation}
    \mathcal{N}(\tau, x) = \frac{1}{2} \Bigg[  1 - \frac{\Delta(\tau, x)}{\sqrt{\Delta(\tau, x)^2+\Gamma(\tau, x)^2+ \Pi(\tau, x)^2}}   \Bigg]
\end{equation}
exceeds zero, i.e., $\mathcal{N}\geq 0$. For convenience, we express all quantities as a function of the dimensionless time, $\tau = \omega_c t$, and the non-Markovianity witness parameter, $x=\tau_R/\tau_E = \omega_c / \omega_0$, which compares the correlation time-scale $\tau_E$ of the environment with the relaxation time-scale $\tau_R$ as well as the ratio of the natural frequency of the probe state, $\omega_0$, with the cutoff frequency of the environment, $\omega_c$. This quantity embodies the rate at which the system's state changes due to its interaction with the environment, running like a witness for the Non-Markovian regime.  In turn, the Markovian regime is attained in the limit, $x \rightarrow \infty$ ensuring $\mathcal{N} \rightarrow 0$ for all times~\cite{IlluminatiPRA2018}. It is important to note that certain approximations, such as the rotating wave approximation and/or the secular approximation, modify the structure or dependence of these coefficients. This affects the proper non-Markovian evaluation, leading to the system being effectively described by Markovian behavior in these regimes, with $\mathcal{N}=0$~\cite{VasilePRA2009,IlluminatiPRA2018}.

\subsection{\texorpdfstring{The Quantum Fisher Information (QFI) \( \mathcal{F}_\theta \)}{The Quantum Fisher Information (QFI) Fθ}}

The Quantum Fisher Information (QFI) quantifies the amount of information that a quantum state carries about an unknown parameter $\theta $, and in this context, the Quantum Cramér-Rao Bound (QCRB) \cite{GiovannettiPRL2006} establishes the lower bound on the standard
deviation $\Delta \theta=\sqrt{\langle \theta^2 \rangle-\langle \theta \rangle^2}$ of the estimated parameter, 
\begin{equation}\label{EQ_CRI}
   \Delta \theta \geq \frac{1}{\sqrt{n \mathcal{F}_{\theta}}} ,
\end{equation}
where $n$ denotes the number of times the experiment is conducted. Hence, this quantity dictates the achievable accuracy of the estimated value, serving as the figure of merit in parameter estimation problems. It is important to note that this inequality holds for unbiased estimators, i.e., estimators for which $\langle \theta_{est} \rangle = \theta_{real}$. Generally, calculating the QFI is a complex task, as it requires the maximization over all Positive Operator-Valued Measures (POVMs)~\cite{serafini2017quantum}. Nevertheless, for the case of a single-mode Gaussian state, with covariance matrix $\boldsymbol{\sigma}$ and first moments $\boldsymbol{d}$, the QFI for estimation of the parameter $\theta$ can be explicitly determined as~\cite{serafini2017quantum,monras2013ARXIV,PinelPRA2013,Jonas_2024,Jonas_2025}

\begin{equation}\label{eq:fisher_serafini}
\mathcal{F}_{\theta}=\frac{\text{Tr}[(\boldsymbol{\sigma^{-1}}\partial_{\theta}\boldsymbol{\sigma})^{2}]}{2(1+\mu^{2})}+2\frac{(\partial_{\theta}\mu)^2}{1-\mu^4}+2(\partial_{\theta}\boldsymbol{d})^{\text{T}}(\boldsymbol{\sigma^{-1}})(\partial_{\theta}\boldsymbol{d}),  
\end{equation}
where $\mu = 1/(2\sqrt{\text{det}(\boldsymbol{\sigma})})$ is the purity of the quantum state and $\partial_{\theta}$ represents the derivative with respect to the parameter $\theta$. The first term is associated with the dynamical dependence of the covariance matrix on the parameter $\theta$. The second one is the dynamic of purity under $\theta$ variation, and the third is the contribution of the moment dynamics of the Gaussian state for the estimated parameter.

Here, we look into how the evolution under the non-Markovian regime can offer a metrological advantage relative to the scenarios extensively explored within the framework of Born-Markov approximations. Interestingly, the QFI flow $d\mathcal{F}_{\theta}/dt$ can serve as a statistical tool to distinguish between Markovian and non-Markovian processes~\cite{LuPRA2010}, and these signatures can be observed in the results presented below. In the following, we analyze quantum thermometry and the metrology of the non-Markovianity witness parameter, setting $\theta=T$, $\theta=x$  in Eq. (\ref{eq:fisher_serafini}), respectively. Moreover, for completeness, we show in Appendix~\ref{Ap: PM} the estimation of PM correlation under the non-Markovian evolution.

\section{Results}\label{sec:re}

\subsection{Non-Markovianity Quantifiers}

\begin{figure}[h]
\centering
\includegraphics[scale = 0.27]{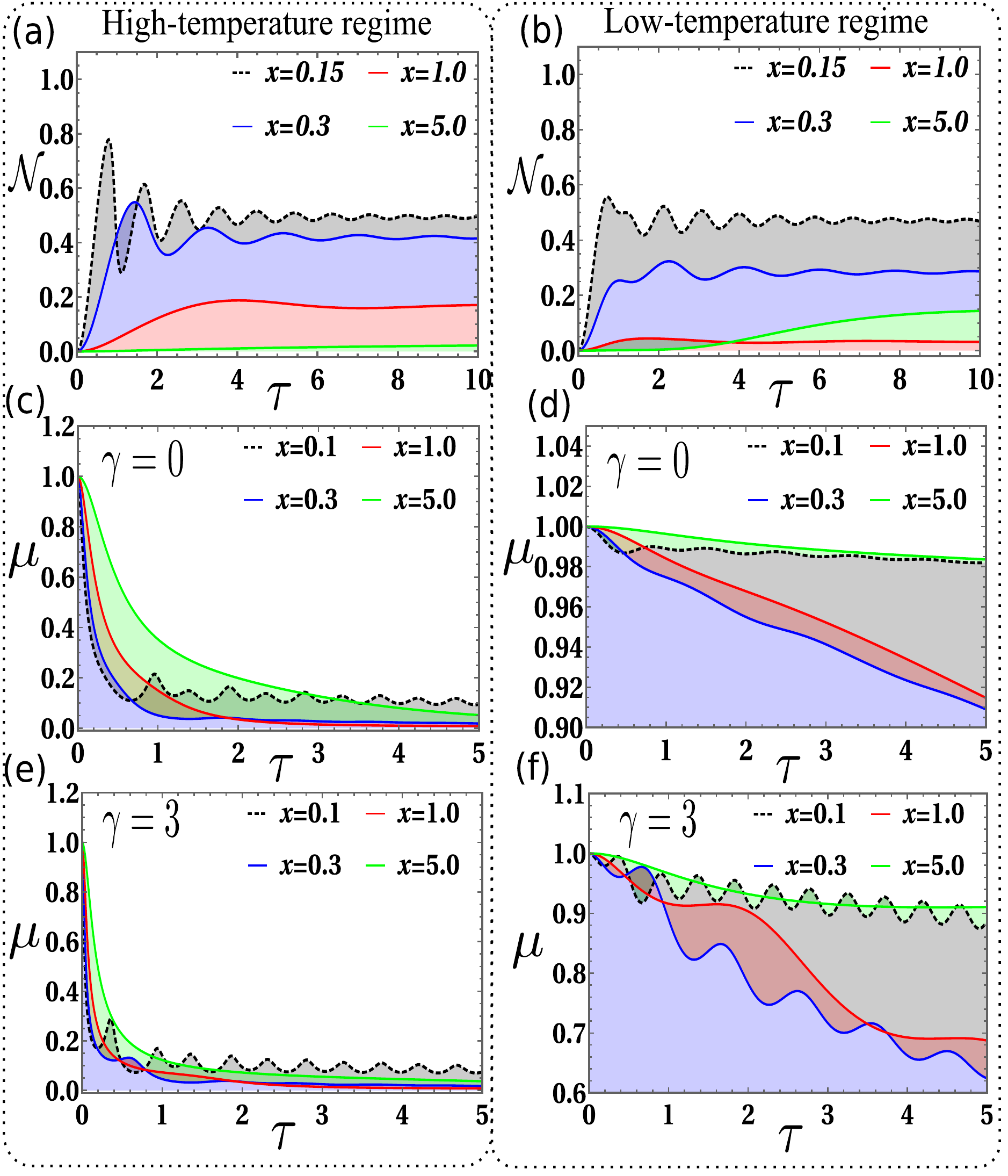}
\caption{Temporal behavior of the non-Markovianity quantifier $\mathcal{N}$  and purity $\mu$ (with different initial PM correlations $\gamma$) under high-temperature 
($k_{B}T/\hbar\omega_{c}=1000$, left panels) and low-temperature ($k_{B}T/\hbar\omega_{c}=10$, right panels) regimes. Even in an almost Markovian regime ($x=5.0$), the non-Markovian characteristic persists only in the $\mathcal{N}$ plots. This contrasts with the revival signatures, which are typically observed only in the short-time dynamics in most cases, particularly when state-dependent quantifiers of non-Markovianity, such as trace distance or purity, are used. Additionally, we observe that the initial PM correlation $\gamma$ enhances the oscillations and revival features of purity compared to the uncorrelated case $\gamma=0$, emphasizing its non-Markovian nature.}
\label{fig2}
\end{figure}

To properly quantify metrology in the non-Markovian regime, we compare two approaches for describing memory effects in an open quantum system. The first method relies on quantifiers that characterize the system's state evolution under a non-Markovian channel, such as trace distance \cite{PiiloPRL2009}, fidelity or Bures distance \cite{BreuerPRA2020}, and purity or entanglement~\cite{FrancoIJQI2016}. For simplicity, we focus on purity in our analysis.  

In this context, Markovian behavior is defined by a monotonic decrease in the quantumness of the system, accompanied by a continuous loss of information from the open quantum system to the environment \cite{BreuerRevModPhysics2016}. A key signature of non-Markovian memory effects is the revival of genuine quantum properties, such as quantum coherence or purity, along with an information backflow from the environment \cite{BreuerRevModPhysics2016}. However, these methods provide only a sufficient, though not necessary, condition to identify non-Markovian behavior in a quantum channel \cite{Rivas_2014}.  For completeness, we also introduce a well-known non-Markovianity quantifier that depends solely on the channel parameters \cite{IlluminatiPRL2015,IlluminatiPRA2018}. This measurement is based on violating the dynamic map's divisibility property.

In Fig. \ref{fig2}, we show the temporal evolution of the non-Markovianity quantifier $\mathcal{N}$ and purity $\mu$ with different initial PM correlations $\gamma$ under high-temperature (left panels) and low-temperature regimes (right panels). Even in an almost Markovian regime ($x=5.0$), the non-Markovian behavior remains visible only in the results of $\mathcal{N}$ for both the high and low-temperature regimes. Furthermore, the initial PM correlation $\gamma$ amplifies the oscillations and revival features of purity compared to the uncorrelated case ($\gamma=0$), further emphasizing its non-Markovian nature.  Additionally, we choose the dimensionless system-reservoir coupling constant as $\alpha = 0.1$, ensuring operation in the weak-coupling regime (see Appendix \ref{Ap:QBM}).  In the following, we analyze the QFI in a non-Markovian scenario and its transition to the conventional Markovian case.

\subsection{Quantum Thermometry}

\begin{figure}[ht]
\centering
\includegraphics[scale = 0.31]{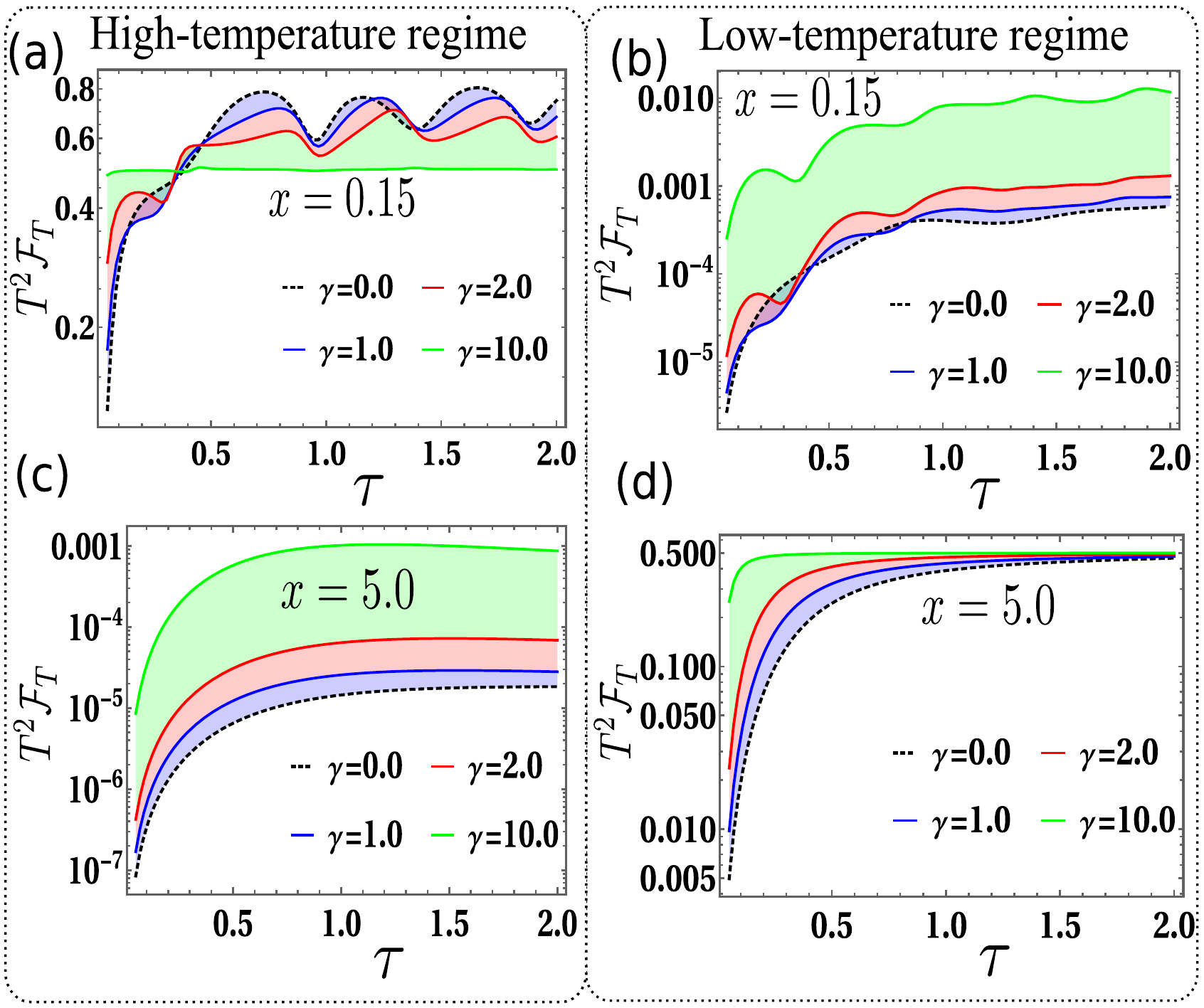}
\caption{The temporal evolution of the Fisher information $T^2\mathcal{F}_T$ (varying the initial PM correlations $\gamma$) is displayed under both high-temperature ($k_{B}T/\hbar\omega_{c}=1000$, left panels) and low-temperature regimes ($k_{B}T/\hbar\omega_{c}=10$, right panels) and for different non-Markovianity witness parameter $x$.  The Fisher information $\mathcal{I}_0 = \mathcal{F}_T(\tau, x, T, \gamma = 0)$ for the uncorrelated probe ($\gamma = 0$) represents the shot-noise limit. }
\label{fig3}
\end{figure}

We start by analyzing the thermometry setup $\theta=T$  in Eq. (\ref{eq:fisher_serafini}). Figure \ref{fig3} illustrates the temporal evolution of the QFI,  $T^2\mathcal{F}_T$, with different initial PM correlations $\gamma$, considering both high-temperature (left panels) and low-temperature (right panels) regimes. The analysis is performed for various values of the non-Markovianity witness parameter $x$. The Fisher information $\mathcal{I}_0 = \mathcal{F}_T(\tau, x, T, \gamma = 0)$, evaluated for the uncorrelated state ($\gamma = 0$), serves as the reference corresponding to the shot-noise limit (SNL). Interestingly, we observe that, in most situations, the PM correlation $\gamma$ works as a valuable resource for metrological applications. This holds across both high- and low-temperature regimes and even in a nearly non-Markovian channel ($x=5.0$). In most cases, we have verified that the accuracy of the measurements improves as the absolute value $\gamma$ increases, with negative values of $\gamma$ yielding results that are not significantly different from those obtained with positive values. Conversely, at high temperatures, as illustrated in Figs.~\ref{fig3}(a) and ~\ref{fig3}(c), the time window during which the advantage of PM correlations can be exploited for more precise temperature estimation becomes significantly narrower. We attribute this effect to a reduction in coherence time, caused by the increased influence of thermal noise under high-temperature conditions. As a result, the system rapidly thermalizes, and its behavior becomes nearly indistinguishable regardless of whether it was initially prepared in a PM-correlated state or not.

\subsection{Non-Markovianity Estimation}

Here, we present what we believe to be the first estimation of the non-Markovianity witness parameter of a quantum channel, encoded by the $x$ parameter. The analysis is done through the QFI, setting $\theta = x$ in Eq. (\ref{eq:fisher_serafini}), and investigating its properties under a high- and low-temperature domain. Specifically, we quantify the metrological enhancement due to the non-Markovian dynamics and study the role of PM correlation $\gamma$ in this regime.  In what follows, we compare the quantum Fisher information (QFI) $\mathcal{F}_x$ for a correlated state $\gamma\neq 0$, with that of an initially uncorrelated Gaussian state $(\gamma = 0)$. The uncorrelated state has uncertainties identical to the vacuum state, establishing the shot noise limit (SNL). This implies that when the probe exhibits PM correlations ($\gamma\neq 0$), a metrological advantage is achieved beyond the shot-noise limit.

\begin{figure}[h]
\centering
\includegraphics[scale = 0.28]{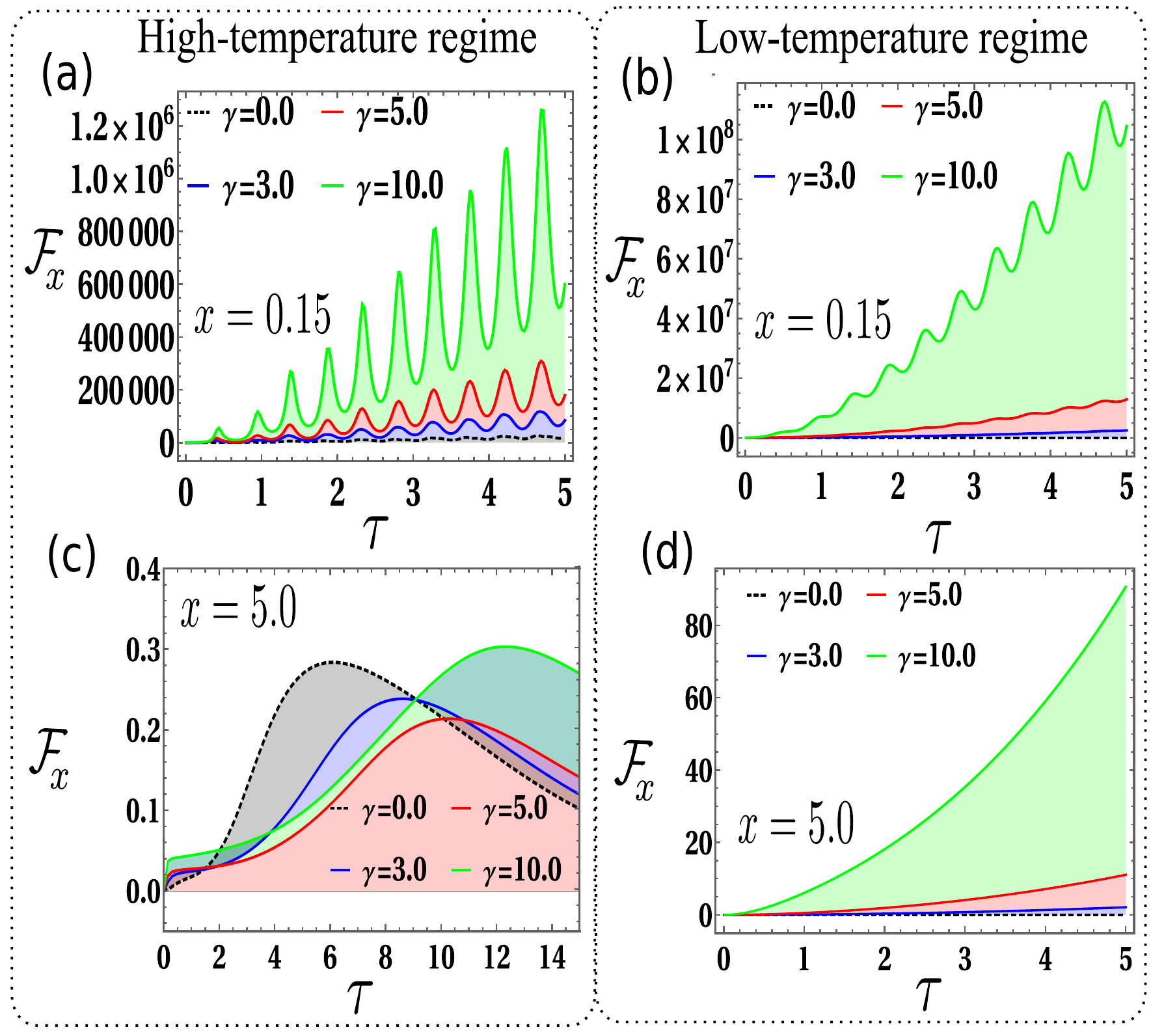}
\caption{The temporal behavior of the Quantum Fisher Information (QFI) $\mathcal{F}_x$  (varying the initial PM correlations $\gamma$) is displayed under both high-temperature ($k_{B}T/\hbar\omega_{c}=1000$, left panels) and low-temperature regimes ($k_{B}T/\hbar\omega_{c}=10$, right panels). Additionally, we analyze the QFI $\mathcal{F}_x$ under a high (low) non-Markovian environment in the top (bottom) panels. The QFI $\mathcal{F}_x$ of an initially uncorrelated Gaussian state $(\gamma = 0)$ corresponds to the SNL. Observe a meteorological enhancement (beating the SNL) as the PM correlation $\gamma$ and the non-Markovianity quantifier (small values of $x$) increase.}
\label{fig4}
\end{figure}

\begin{figure}[h]
\centering
\includegraphics[scale = 0.31]{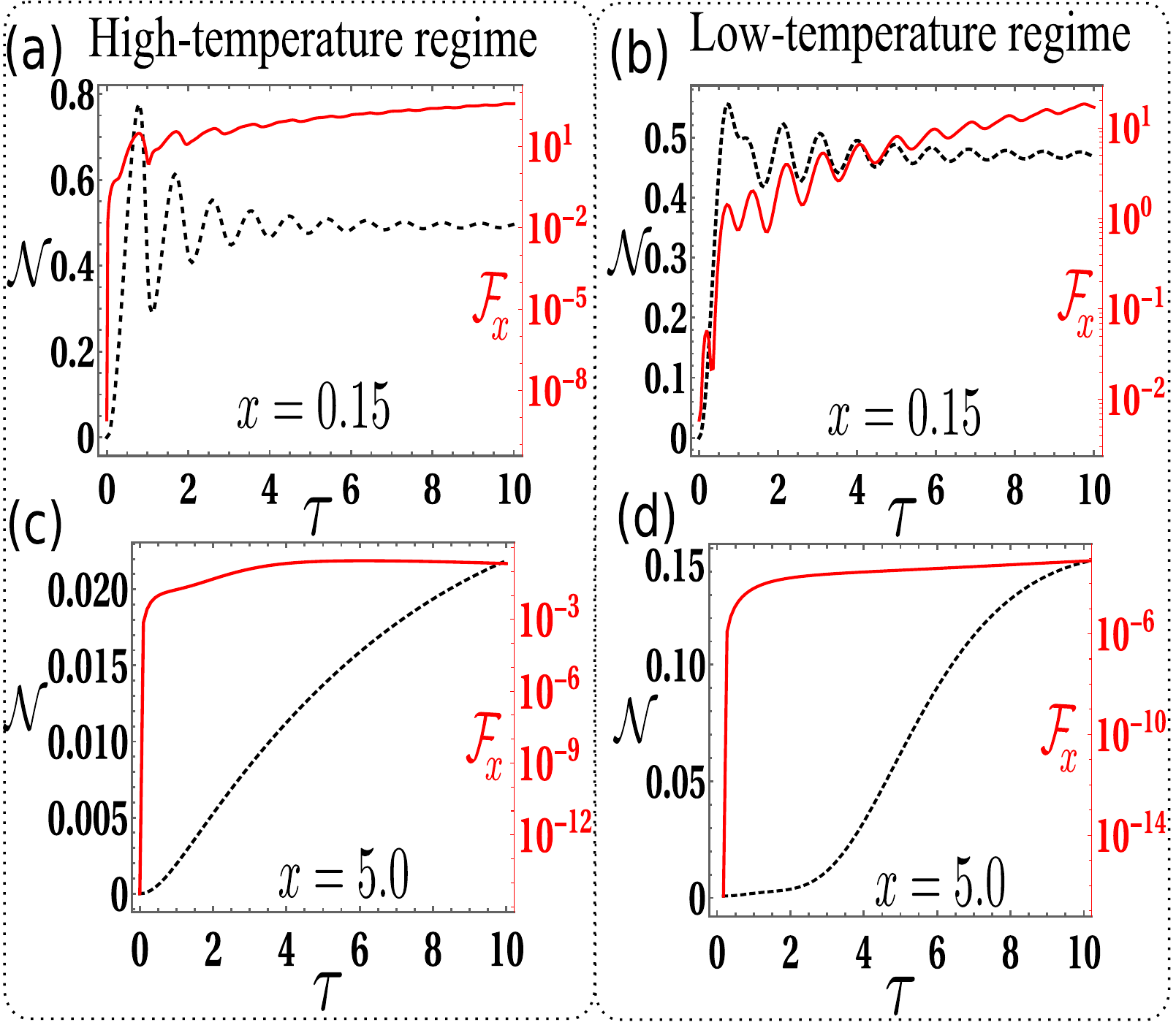}
\caption{Dynamic of the QFI, $\mathcal{F}^{\gamma=0}_x$ (solid red line), and non-Markovianity quantifier, $\mathcal{N}$ (black dashed line), for both high-temperature ($k_{B}T/\hbar\omega_{c}=1000$) and low-temperature conditions ($k_{B}T/\hbar\omega_{c}=10$) considering  different non-Markovianity witness parameter $x$.}
\label{fig5}
\end{figure}

Figure \ref{fig4} illustrates the temporal evolution of QFI $\mathcal{F}_x$ for different values of the initial PM correlations in high-temperature (left panels) and low-temperature regimes (right panels). Furthermore, we analyze the QFI $\mathcal{F}_x$ in a high (low) non-Markovian environment in the top (bottom) panels. A metrological improvement (which exceeds the SNL) is observed as the PM correlation $\gamma$ and the non-Markovianity quantifier (small values of $x$) increase. Furthermore, the results demonstrate how PM correlation can improve metrological performance across various configurations. Interestingly, for regimes with high values of $x$ and high temperatures [Fig.~\ref{fig4}(c)], we observe a temporal transition from a gain to a loss region in metrological performance, followed by a return to a gain region. We interpret this behavior as a consequence of a backflow of information from the environment to the system, which still occurs in this scenario but more slowly and gradually. This contrasts with regimes characterized by small values of $x$ (non-Markovian), where the exchange of information between the system and environment is faster. This suggests that, at later times of the dynamics, the initial correlation enhances the estimation of the non-Markovianity parameter $x$, an effect that persists even under high-temperature conditions and short environmental correlation times ($\tau_E \ll 1$ or $x \gg 1$). This connection is examined in more detail in Fig.~\ref{fig5}.

To better understand the relationship between the quantum Fisher information (QFI) $\mathcal{F}_x$ and the non-Markovianity quantifier $\mathcal{N}$, we present in Fig. \ref{fig5} their evolution under different physical conditions. The dashed black line represents $\mathcal{N}$, while the solid red line corresponds to $\mathcal{F}_x$. Interestingly, in a highly non-Markovian regime ($x=0.15$), the revival oscillations are associated with an increase in QFI $\mathcal{F}^{\gamma=0}_x$, which can be associated with the backflow dynamic of information from the environment. As expected, in the nearly Markovian limit ($x=5.0$), the oscillatory revivals vanish, and the corresponding QFI increases only monotonically. However, the information gain associated with the backflow remains present, though it now increases at a slower and smaller rate compared to the non-Markovian case.

\section{Discussion}\label{sec:disc}

Understanding quantum metrology in non-Markovian environments is crucial for advancing quantum technologies beyond the limitations of the Born-Markov approximation. This work investigates how non-Markovian evolution can provide a metrological advantage over conventional Markovian scenarios. Through the study of quantum thermometry, we demonstrate that non-Markovian dynamics enhance parameter estimation, offering valuable insights for applications such as quantum thermodynamics. Additionally, we introduce a method for estimating non-Markovianity across different temperature regimes and show how initial position-momentum (PM) correlations can improve this estimation, thereby extending the results of Ref.~\cite{Porto2024} to the non-Markovian domain.

Our findings highlight the role of initial PM correlations and the non-Markovian witness parameter $x$ in metrological performance. When $\gamma \neq 0$, PM correlations amplify oscillations and revivals in purity, reinforcing their link to non-Markovian behavior. The temporal evolution of the quantum Fisher information (QFI) $\mathcal{F}_T$ under different initial PM correlations and temperature regimes reveals that PM correlation enhances metrological precision in most cases, even in weakly non-Markovian channels. Furthermore, the analysis of metrological gain as a function of $\gamma$ and $x$ confirms that PM correlations significantly improve estimation accuracy, particularly in strongly non-Markovian environments, where the gain surpasses the shot-noise limit (SNL).

Future research could extend these findings by exploring alternative spectral densities beyond the Ohmic-like model to assess their impact on metrological gain. Additionally, experimental implementations in optomechanical or superconducting systems would provide a valuable test of these theoretical predictions, bringing practical quantum sensing applications closer to realization.

\begin{acknowledgments}
J.C.P.P. acknowledges Fundação de Amparo à Pesquisa do Estado do Piauí (FAPEPI) for financial support. C.H.S.V. acknowledges the São Paulo Research Foundation (FAPESP), Grant No. 2023/13362-0, for financial support and the Federal University of ABC (UFABC) for providing the workspace. I.G.P. acknowledges Grant No. 306528/2023-1 from CNPq. P.R.D. acknowledges support from the NCN Poland, ChistEra-2023/05/Y/ST2/00005 under the project Modern Device Independent Cryptography (MoDIC). L.S.M. acknowledges support from the National Institute of Science and Technology on National Institute of Photonics (INFO) CNPq - INCT grant 409174/2024-6.
\end{acknowledgments}

\appendix

\section{QUANTUM BROWNIAN
MOTION}\label{Ap:QBM}

A brief overview, based on Refs. \cite{IlluminatiPRA2018,VasilePRA2009}, of the QBM exact master equation and its solutions is provided here. This model describes the evolution of an open quantum-mechanical oscillator system, characterized by a frequency $\omega_0$, linearly coupled to an infinite bath (environment) of harmonic oscillators. The global Hamiltonian is given by \cite{Petruccione2002book,BreuerPRA2020}
\begin{gather}
    H = \underbrace{ \frac{\hat{p}^2}{2m}+\frac{1}{2}m\omega_0\hat{q}^2}_{H_S} \;+ \; \underbrace{\sum_n \Big( \frac{\hat{p}_n^2}{2m_n}+\frac{1}{2}m_n\omega_n\hat{q}^2_n \Big)}_{H_E}
    \; \nonumber \\ - \;\underbrace{\hat{q}\sum_n \kappa_n \hat{q}_n}_{H_I}  \;+ \;\underbrace{\hat{q}^2\sum_n \frac{\kappa_n^2}{2m_n \omega_n^2}}_{H_C}.
\end{gather}

Here, $\hat{q}$ and $\hat{p}$ are position and momentum operators of the central oscillator, while the $\hat{q}_n$ and $\hat{p}_n$ are the corresponding bath operators. $H_S$ and $H_E$ are the self and environmental Hamiltonians, respectively. $H_I$ corresponds to the interaction part, where $\kappa_n$ represents the position coupling strength between the system and the $n$th
bath oscillator. The system is assumed to interact with the bath through linear coupling to a collection of harmonic modes \cite{FEYNMAN_Vernon1963}. The last term is introduced to compensate for the small renormalization of the central frequency $\omega_0$ due to interaction with the bath. The associated exact master equation describing the reduced density matrix $\rho$ of the system, under the particular case
of a factorized initial system-bath state, is given by \cite{JPPaz1992PRD,IlluminatiPRA2018}
\begin{gather}
 \frac{d \rho (t)}{dt}  = - \frac{i}{\hbar} [H_S,\rho (t)] - \Delta(t) [\hat{q}, [\hat{q},\rho (t)]] + \Pi(t) [\hat{q}, [\hat{p},\rho (t)]] \nonumber \\ - i\Gamma(t) [\hat{q}, [\hat{p},\rho (t)]],\label{eq:ExactMasterEquation}
\end{gather}
where $\Gamma(t)$ is the damping coefficient, $\Delta(t)$ and  $\Pi(t)$ are the normal and anomalous diffusion coefficients, respectively. This master equation local in time could appear surprising since in general non-Markovian systems are described by a nonlocal
memory kernel taking into account the history of the reduced system \cite{intravaia2003PRA}. However, the benefit of this equation is that the memory
effects of the environment are incorporated into its time-dependent coefficients. This characteristic is inherent in all generalized master equations obtained through the time-convolutionless projection operator method \cite{Petruccione2002book}. Also, since neither the Born-Markov approximation nor the secular
approximation has been performed, this equation is considered exact \cite{VasilePRA2009}. This equation can be solved exactly, although the time-dependent coefficients have no obvious closed form \cite{JPPaz1992PRD}; therefore, it is necessary to resort to numerical methods. Even so, this is a very difficult task, and in general, the dynamics of the system are analyzed only in the weak-coupling limit, which preserves the form of the master equation for the dynamical map, but does not invoke the Born-Markov and rotating wave approximations, and is given by \cite{intravaia2003PRA}
\begin{equation}
    \Gamma (t) = \alpha^2
    \int_{0}^{t} \int_{0}^{+ \infty} 
 d t' d\omega J(\omega) \sin(\omega t') \sin(\omega_0 t'),
\end{equation}
\begin{gather}
    \Delta (t)= \alpha^2  \int_{0}^{t} \int_{0}^{+ \infty}  d t' d\omega J(\omega) [2N(\omega)+1] \nonumber \\
    \times \cos(\omega t') \cos(\omega_0 t'),
\end{gather}
and
\begin{gather}
    \Pi (t) = \alpha^2  \int_{0}^{t} \int_{0}^{+ \infty}  d t' d\omega J(\omega) [2N(\omega)+1] \nonumber\\
    \times \cos(\omega t') \sin(\omega_0 t'),
\end{gather}
where $\alpha$ is the dimensionless
system-reservoir coupling constant, $N(\omega)=[\exp(\hbar\omega/k_B T)-1]^{-1}$ is the thermal occupation number, and $J(\omega) = \sum_n  \kappa_n^2/ (2 m_n \omega_n^2 ) \delta (\omega - \omega_n)$ is the spectral density that models the system-environment interaction and determines the bath characteristics, with $\omega$ the bath oscillator's frequency. In the continuum limit, this function is phenomenologically represented by a smooth function of $\omega$ [see Eq. (\ref{eq:spectral_density})].

Closed expressions of the time-dependent coefficients of the master equation (\ref{eq:ExactMasterEquation}) can be obtained in the high and low-temperature regime, i.e., for $[2N(\omega)+1] =\coth \Big( \frac{\hbar \omega}{2 k_B T} \Big) \approx \frac{2 k_B T }{\hbar \omega} $ and $[2N(\omega)+1] = \coth \Big( \frac{\hbar \omega}{2 k_B T} \Big) \approx 1 +2\exp{\Big( -\frac{\hbar \omega}{k_B T} \Big)}$, respectively (for more details, see the Appendix of Ref. \cite{VasilePRA2009}). These are the cases analyzed here and are written explicitly in the Appendix \ref{Ap:Coeff}. In general, after a time $t \approx \omega_c^{-1}$, the coefficients attain their Markovian stationary values and the system behaves according to the predictions of the Markovian theory \cite{VasilePRA2009}.

We restrict ourselves to solving the dynamics of Gaussian states, which have a simple Gaussian characteristic function characterized by its first-order moments $\boldsymbol{d}$ and its covariance matrix $\boldsymbol{\sigma}$ \cite{serafini2017quantum}. In particular, the evolution of these quantities through an $N$-mode Gaussian quantum channel (environment) can be given by \cite{VasilePRA2009,IlluminatiPRA2018,PolzikBook2007}
\begin{equation}
   \boldsymbol{d}(t) = \mathcal{S}(t) \boldsymbol{d}(0) \; , \;\;\;\;\;\;\;\;\;\;\;  \boldsymbol{\sigma} (t) = \mathcal{S}(t) \boldsymbol{\sigma}(0) \mathcal{S}(t)^{\mathsf{T}} +\mathcal{T}(t),
\end{equation}
where $\mathcal{S}(t)$ and $ \mathcal{T}(t)$ are $2N \times 2N$ matrices that characterize the exact
evolution of Eq. (\ref{eq:ExactMasterEquation}) and given by \cite{IlluminatiPRA2018}
\begin{gather}
    \mathcal{S}(t) = e^{-[\widetilde{\Gamma} (t)/2]} R(t), \;\;\;\;\;\;\;\;\;\;\;\;\;\;\; \text{and} \nonumber \\  \mathcal{T}(t) =  [R^{-1}(t)]^{\mathsf{T}}\Big[e^{-\widetilde{\Gamma} (t)} \int_{0}^{t} d t' e^{\widetilde{\Gamma} (t')}  R^{\mathsf{T}}(t') M(t') R(t')    \Big]  R^{-1}(t), \label{eq:X_Y}
\end{gather}
with
\begin{gather}\label{eq:M_R_matrices}
M(t)=\left(\begin{array}{cc}
\Delta(t) & -\Pi(t)/2  \\
-\Pi(t)/2 & 0
\end{array}\right), \nonumber \\  R(t)=\left(\begin{array}{cc}
\cos (\omega_0 t) & \sin (\omega_0 t)  \\
-\sin (\omega_0 t)   & \cos (\omega_0 t)
\end{array}\right),   \;\;\;\;\; \widetilde{\Gamma} (t) = 2 \int_{0}^{t} \Gamma (t') dt'.
\end{gather}

We are especially focused on the short-time dynamics that are non-Markovian; then, in this regime and within the weak coupling limit, we can expand the bracket term appearing in (\ref{eq:X_Y}) as

\begin{gather}
    \Big[e^{-\widetilde{\Gamma}(t)} \int_{0}^{t} d t' e^{\widetilde{\Gamma} (t')}  R^{\mathsf{T}}(t') M(t') R(t')    \Big]  \approx \nonumber \\ \int_{0}^{t} d t'  R^{\mathsf{T}}(t') M(t')  R(t')   - \widetilde{\Gamma}(t) \int_{0}^{t} d t'  R^{\mathsf{T}}(t') M(t')R(t') \nonumber \\ +  \int_{0}^{t} d t' \widetilde{\Gamma}(t') R^{\mathsf{T}}(t')  M(t')R(t')  + \mathcal{O}(\alpha^4).
\end{gather}
Once $\widetilde{\Gamma}(t)  \propto \alpha^2$ and $M(t) \propto \alpha^2 $, in the weak coupling limit $(\alpha \ll 1)$ and for short non-Markovian times, the first term predominates, making it the only term that will be considered \cite{VasilePRA2009}.

\section{TIME DEPENDENT COEFFICIENTS}\label{Ap:Coeff}

\begin{widetext}
Here, we provide analytic expressions of the time-dependent coefficients of the master equation (\ref{eq:ExactMasterEquation}). For simplicity, we examine only the Ohmic reservoir spectral density ($s=1$) and calculate the temperature-independent damping coefficient $\Gamma(t)$, the diffusion coefficients in the high-temperature regime $\Delta_{T_{High}} (t)$ and $\Pi_{T_{High}} (t)$, and the diffusion coefficients in the low-temperature, $\Delta_{T_{Low}} (t)$ and $\Pi_{T_{Low}} (t)$. In the following, we write these results

\begin{equation}
    \Gamma(\tau,x) = \frac{\alpha^2}{4 x} \Bigg\{ i e^{-1/x} \Big[ \text{Ei} \Big ( \frac{1-i\tau}{x} \Big) - \text{Ei} \Big ( \frac{1+i\tau}{x} \Big) \Big] +   e^{1/x} \Big[ 2\pi + i \text{Ei} \Big ( \frac{i\tau-1}{x} \Big) - i \text{Ei} \Big ( \frac{1+i\tau}{x} \Big) \Big] - \frac{4 x\sin (\tau/x)}{1+\tau^2} \Bigg\}, 
\end{equation}
\begin{equation}
    \Delta_{T_{High}} (t,x,T) = \frac{\alpha^2 k_B T e^{-1/x}}{2\hbar \omega_c} \Bigg\{  i \Big [ \text{Ei} \Big ( \frac{1- i\tau}{x} \Big) - \text{Ei} \Big ( \frac{1+ i\tau}{x}\Big)   \Big]  + e^{2/x}\Big[  2\pi +i \text{Ei} \Big ( \frac{i\tau-1}{x}\Big) - i \text{Ei} \Big ( - \frac{i\tau+1}{x}\Big)  \Big]   \Bigg\},
\end{equation}
\begin{equation}
    \Pi_{T_{High}}(t,x,T) = \frac{\alpha^2 k_B T e^{-1/x}}{2\hbar \omega_c} \Bigg\{   -\text{Ei} \Big ( \frac{1- i\tau}{x} \Big) - \text{Ei} \Big ( \frac{1+ i\tau}{x}\Big) + 2\text{Ei} \Big (\frac{1}{x}\Big)     + e^{2/x}\Big[ -2\text{Ei} \Big (-\frac{1}{x}\Big)  +\text{Ei} \Big ( \frac{i\tau-1}{x}\Big) + \text{Ei} \Big ( - \frac{i\tau+1}{x}\Big)  \Big]   \Bigg\},
\end{equation}
\begin{gather}
    \Delta_{T_{Low}}(t,x,T) = \frac{\alpha^2}{4x} \Bigg \{  \frac{4 x \cos (\tau/x)}{1+\tau^2} + ie^{-1/x} \Big[\text{Ei}\Big(\frac{1-i\tau}{x}\Big) -\text{Ei}\Big(\frac{1+i\tau}{x}\Big)  \Big] - e^{1/x} \Big[ 2\pi +i\text{Ei}\Big(\frac{-1+i\tau}{x}\Big) -i\text{Ei}\Big(-\frac{1+i\tau}{x}\Big)  \Big] \Bigg\} \nonumber \\
    +\frac{2\alpha^2\tau\cos(\tau/x)}{\tau^2+[1+(\hbar\omega_c)/(k_BT)]^2} 
    +\frac{\alpha^2}{x} \Bigg\{ \Bigg [i \text{Ci}\Bigg(\frac{\tau-i[1+(\hbar\omega_c)/(k_BT)]}{x}  \Bigg)-i \text{Ci}\Bigg(\frac{\tau+i[1+(\hbar\omega_c)/(k_BT)]}{x}  \Bigg) -\pi \Bigg]\nonumber \\  \times\sinh{\Bigg[ \frac{1+(\hbar\omega_c)/(k_BT)}{x} \Bigg]} 
    +\cosh{\Bigg[ \frac{1+(\hbar\omega_c)/(k_BT)}{x} \Bigg]} \Bigg [ \text{Si}\Bigg(\frac{\tau-i[1+(\hbar\omega_c)/(k_BT)]}{x}  \Bigg)+ \text{Si}\Bigg(\frac{\tau+i[1+(\hbar\omega_c)/(k_BT)]}{x}  \Bigg) \Bigg]
 \Bigg\},
\end{gather}
and
\begin{gather}
    \Pi_{T_{Low}}(t,x,T) = \frac{\alpha^2}{4x} \Bigg \{  \frac{4 x \sin (\tau/x)}{1+\tau^2} -e^{-1/x} \Big[\text{Ei}\Big(\frac{1-i\tau}{x}\Big) +\text{Ei}\Big(\frac{1+i\tau}{x}\Big) -2\text{Ei}\Big(\frac{1}{x}\Big)  \Big] 
    + e^{1/x} \Big[2\text{Ei}\Big(-\frac{1}{x}\Big) -\text{Ei}\Big(\frac{-1+i\tau}{x}\Big)\nonumber\\ -\text{Ei}\Big(-\frac{1+i\tau}{x}\Big)  \Big] \Bigg\} +\frac{2\alpha^2\tau\sin(\tau/x)}{\tau^2+[1+(\hbar\omega_c)/(k_BT)]^2}
    +\frac{\alpha^2}{x} \Bigg\{ \cosh{\Bigg[ \frac{1+(\hbar\omega_c)/(k_BT)}{x} \Bigg]}\Bigg [ \text{Ci} \Bigg(-i \frac{1+(\hbar\omega_c)/(k_BT)}{x} \Bigg) \nonumber \\ +  \text{Ci} \Bigg(i \frac{1+(\hbar\omega_c)/(k_BT)}{x} \Bigg)  - \text{Ci}\Bigg(\frac{\tau-i[1+(\hbar\omega_c)/(k_BT)]}{x}  \Bigg) - \text{Ci}\Bigg(\frac{\tau+i[1+(\hbar\omega_c)/(k_BT)]}{x}  \Bigg)   \Bigg]  + \nonumber \\
    + \sinh{\Bigg[ \frac{1+(\hbar\omega_c)/(k_BT)}{x} \Bigg]} \Bigg[ -2\text{Shi}\Bigg(\frac{1+(\hbar\omega_c)/(k_BT)}{x}  \Bigg)  +i\text{Si}\Bigg(\frac{\tau-i[1+(\hbar\omega_c)/(k_BT)]}{x}  \Bigg)  -i\text{Si}\Bigg(\frac{\tau+i[1+(\hbar\omega_c)/(k_BT)]}{x}  \Bigg)  \Bigg]
 \Bigg\} ,
    \end{gather}

where
\begin{gather}
    \text{Ei}(z)= -\int_{-z}^{\infty} \frac{e^{-u}}{u} du, \;\; \text{Ci} (z) = -\int_{z}^{\infty} \frac{\cos u}{u} du,\;\;\; \text{Si} (z) = \int_{0}^{z} \frac{\sin u}{u} du,  \;\; \text{Shi} (z) = \int_{0}^{z} \frac{\sinh u}{u} du
\end{gather}
are the exponential integral function, cosine integral function, sine integral function, and hyperbolic sine integral function.

\end{widetext}

\section{How Position-Momentum Correlations Boost Quantum Metrology}\label{Ap:PM_metrology}

In this section, we explore the relationships among position-momentum correlation, squeezing, and Quantum Fisher Information. There is a well-established connection between position-momentum correlations and squeezing~\cite{Marinho_2020,Porto2024}. To illustrate it more explicitly, let us consider a squeezed vacuum state, $|S(\xi)\rangle=S(\xi)|0\rangle$, that has a covariance matrix given by~\cite{LandiPRA2024}:
\begin{equation}\label{eq:squeezed_state}
    \boldsymbol{\sigma} = \frac{1}{2}\begin{pmatrix} \;
\cosh 2r - \sinh 2r \cos\phi &  \sinh 2r \sin\phi  \\
\- \sinh 2r \sin\phi  & \cosh 2r + \sinh 2r \cos\phi \;
\end{pmatrix}.
\end{equation}
Here, $S(\xi)$ is the single-mode squeezing operator with $\xi=re^{i\phi}$, $r$ is the squeezing parameter, and $\phi$ is the phase rotation angle that determines the direction of squeezing in phase space. By comparing the covariance matrix (\ref{eq:squeezed_state}) with Eq. (\ref{eq:iniial_state}), we find that $\gamma = \sinh(2r)\sin\phi$. Therefore, when $\phi=0$, it follows that $\gamma=0$, ensuring that only a squeezed state that has been rotated in phase space exhibits a non-null position-momentum correlation. In contrast, an ordinary coherent state evolving freely with Hamiltonian, $\hat{H}=\hat{p}^2/2m$, under the Schrödinger equation can also acquire position-momentum correlations over time, not as a result of squeezing, but due to its dynamical evolution. In this case, the covariance matrix takes the form \cite{bohm1951quantum}:
\begin{equation}
    \boldsymbol{\sigma} = \frac{1}{2}\begin{pmatrix} \;
1+ \frac{t^2}{\tau_0^2}\;\;\; & \frac{t}{\tau_0}  \\
\frac{t}{\tau_0}  \;\;\;& 1\;
\end{pmatrix},
\end{equation}
where $\tau_0$ is a characteristic evolution time-scale. The key idea is that this type of correlation makes the system’s probability distribution most likely to lie within an elliptical region in phase space, as illustrated in Ref.~\cite{bohm1951quantum}. In simple terms, this can be interpreted as the result of the system's evolution, where large positions become correlated with large momenta, resulting in a stretched, elliptical shape in phase space.

\begin{figure}[h]
\centering
\includegraphics[scale=0.45]{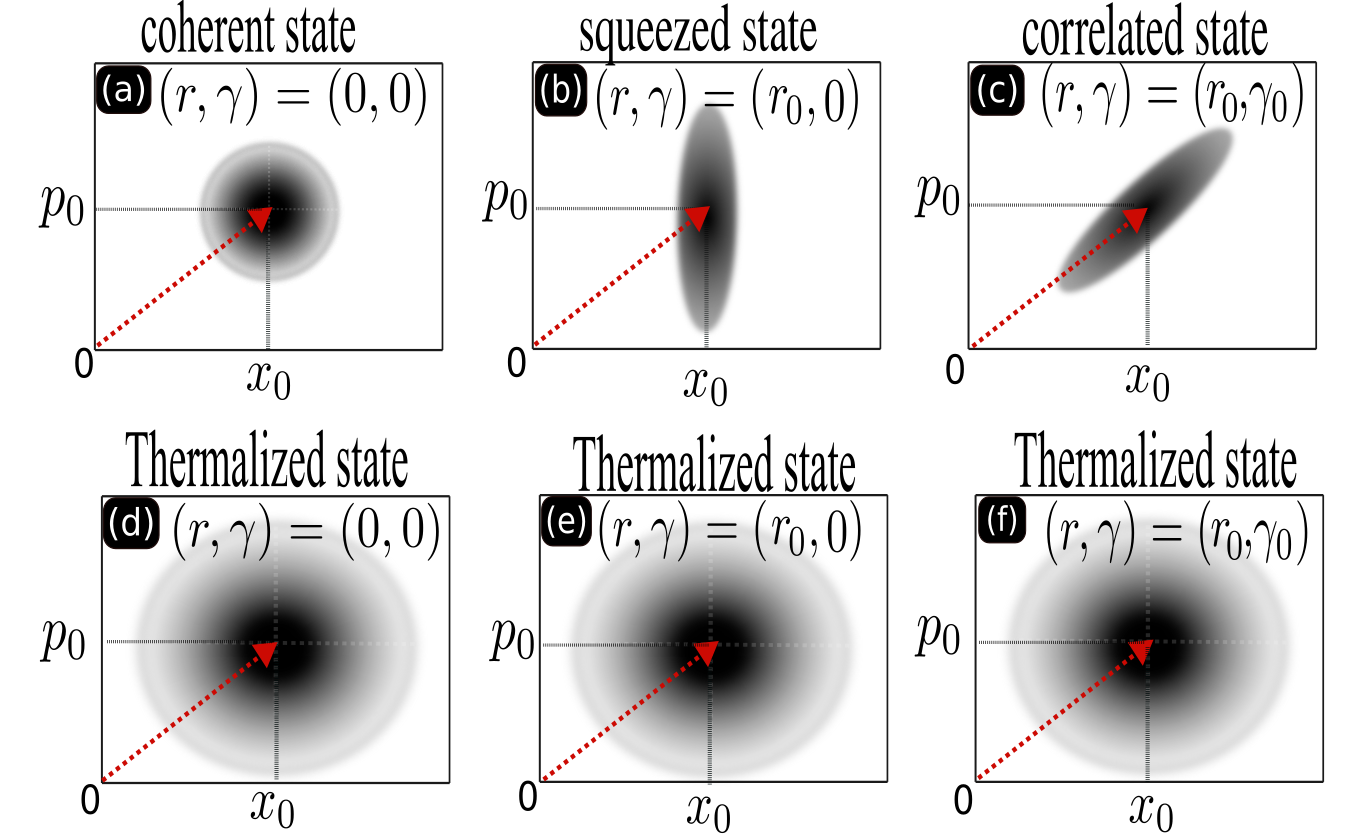}
\caption{Phase-space representation of the Wigner function showing how different initial states evolve through a quantum channel until thermalization. The greater the distinguishability between the initial and final states, the more suitable the state is for metrological applications. 
}\label{figwig}
\end{figure}

To elucidate the role of position-momentum correlation in improving measurement precision, we recall that in parameter estimation theory, the Quantum Fisher Information (QFI) can be seen as a quantifier of distinguishability between a quantum state, $\rho_{\theta}$, and its neighboring state, $\rho_{\theta + \delta\theta}$ \cite{PinelPRA2013}. A greater change in the state in response to variations in the encoded parameter $\theta$ implies higher sensitivity, which in turn corresponds to a larger quantum Fisher information (QFI). By analyzing the evolution of the probability distribution in phase space—using, for instance, the Wigner function—it is possible to note that distinguishability can be enhanced when the initial state exhibits position-momentum correlation. In this case, the distribution forms a rotated, stretched ellipse, which is more sensitive to parameter variations than the circular distribution of an ordinary coherent state, thereby improving metrological precision. The Wigner function plots (see Fig. \ref{figwig}) illustrate this behavior, demonstrating how different initial states evolve under the action of the quantum channel.

\section{PM Correlations Estimation}
\label{Ap: PM}

\begin{figure}[h]
\centering
\includegraphics[scale = 0.3]{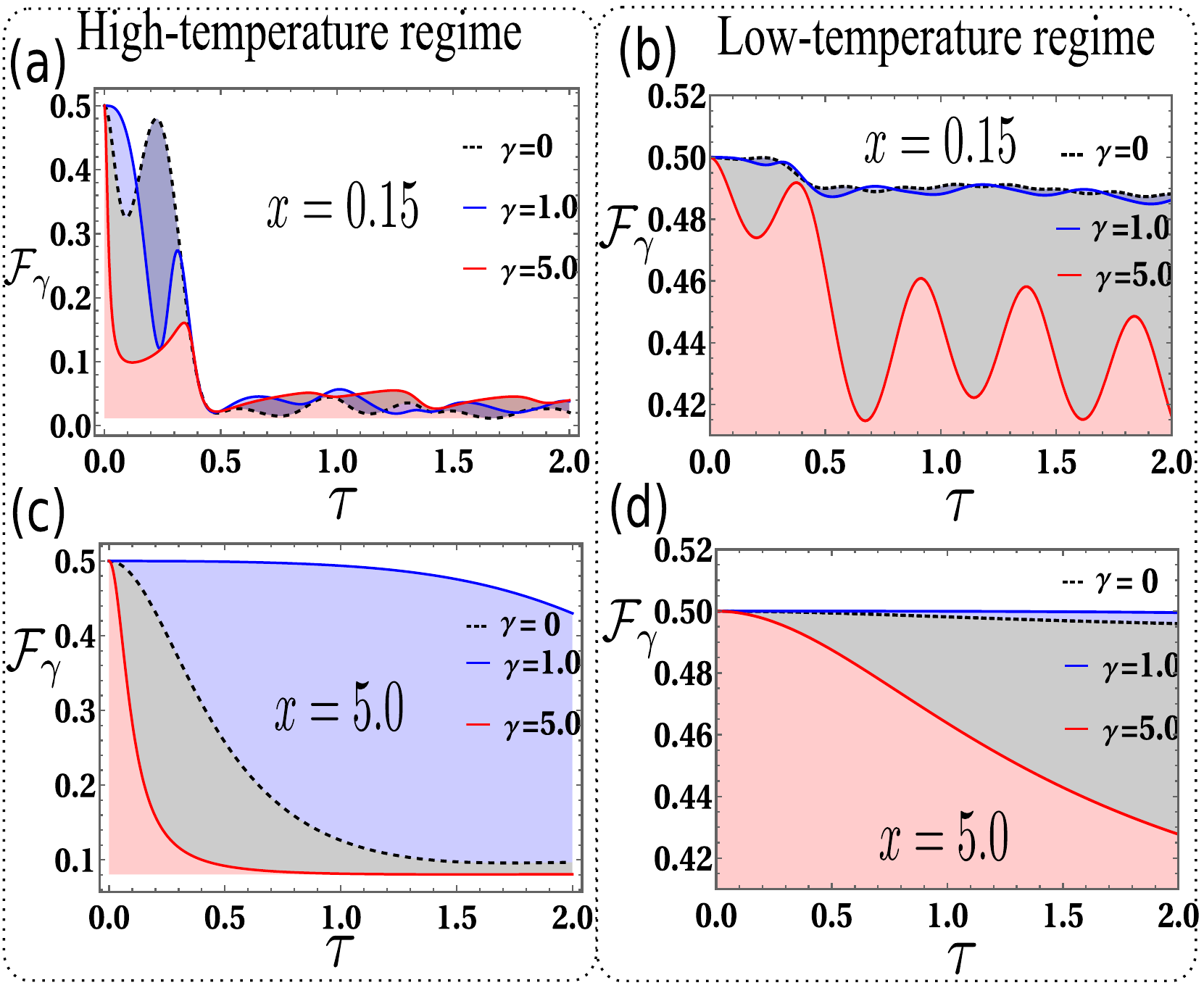}
\caption{The temporal behavior of the QFI $\mathcal{F}_\gamma$ (varying the initial PM correlations $\gamma$) is displayed under both high-temperature ($k_{B}T/\hbar\omega_{c}=1000$ left panels) and low-temperature regimes ($k_{B}T/\hbar\omega_{c}=10$, right panels) and for different non-Markovianity witness parameter $x$. 
}
\label{fig6}
\end{figure}

For the sake of completeness, we present the PM correlation $\gamma$ estimation under various conditions in Fig. \ref{fig6}. The figure shows that, depending on the evolution time $\tau$, an appropriate choice of the initial PM correlation $\gamma$ can maximize the QFI $\mathcal{F}_{\gamma}$.

\FloatBarrier 

\bibliography{references}

\end{document}